\documentclass[10pt,a4paper]{article}
\usepackage{graphics,graphicx}
\usepackage{authblk}
\usepackage{cite}
\usepackage[utf8]{inputenc}

\title{Single-cycle attosecond pulses by Thomson backscattering of terahertz pulses}

\author[1,2,*]{Gy\"orgy T\'oth}
\author[1]{Zolt\'an Tibai}
\author[3]{Ashutosh Sharma}
\author[2,3,4]{J\'ozsef A. F\"ul\"op}
\author[1,2,4]{J\'anos Hebling}

\affil[1]{Institute of physics, University of P\'ecs, 7624 P\'ecs, Hungary}
\affil[2]{MTA-PTE High-Field Terahertz Research Group, 7624 P\'ecs, Hungary}
\affil[3]{ELI-ALPS, ELI-Hu Nkft., 6720 Szeged, Hungary}
\affil[4]{Szent\'agothai Research Centre, University of P\'ecs, 7624 P\'ecs, Hungary}

\begin{document}
\maketitle

\begin{abstract}
The generation of single-cycle attosecond pulses based on Thomson scattering of terahertz (THz) pulses is proposed. In the scheme, a high-quality relativistic electron beam produced by a laser-plasma wakefield accelerator (LPWA), is sent through suitable magnetic devices to produce ultrathin electron layers for coherent Thomson backscattering of intense THz pulses. According to numerical simulations, single-cycle attosecond pulse generation is possible with up to 1 nJ energy. The waveform of the attosecond pulses closely resembles that of the THz pulses. This allows for a flexible waveform control of attosecond pulses.
\end{abstract}

%\setboolean{displaycopyright}{true}

%\begin{document}

%\maketitle
%\thispagestyle{fancy}

%\ifthenelse{\boolean{shortarticle}}{\abscontent}{}

\section{Introduction}

Attosecond pulses can be generated in a number of ways, such as high-harmonic generation \cite{Krausz2009}, undulator radiation \cite{Zholents2004,Saldin2006,Marinelli2013,Dunning2013}, or Thomson scattering \cite{Wu2011,Paz-2012,Lu02013,Hack2017}. High-harmonic generation is capable of producing single-cycle attosecond pulses \cite{Sansone2006}. Even subcycle pulses were generated by laser-generated ultrathin electron layers from double-foil targets \cite{Ma-2014}. However, controlling the pulse shape has not been demonstrated so far. This is also true of radiation sources based on relativistic electrons \cite{Dunning2013,Tanaka2014}. In order to circumwent this limitation, we proposed a technique to generate cerrier-envelope phase stable, waveform-controlled single-cycle \cite{TibaiPRL,TibaiarXive} or few-cycle \cite{Toth2016} attosecond pulses by coherent undulator radiation. In this scheme, ultrathin (of a few nm) relativistic electron layers, generated in a laser-assisted bunching process, emit attosecond pulses when passing through a radiator undulator. The magnetic field distribution of the undulator is transferred to the waveform of the emitted radiation \cite{TibaiPRL,Toth2016,TibaiarXive}, thereby offering a unique waveform tailoring capability. Nevertheless, this technique requires high-energy (GeV) electrons, as an extremely short undulator would be necessary for lower-energy ($<$100 MeV) electrons, which are more easily avaialable even from laser electron accelerators.

A LPWA is able to generate electron beams over cm acceleration distances with parameters comparable to (and in some regards even better than) conventional sources. The pulse duration unique to LPWAs is intrinsically ultrashort, which is more than one order of magnitude shorter than those in X-ray free electron lasers (XFELs). The recent progresses in LPWA \cite{Leemans2006,Lundh-2011,Esarey2009} allow efficient production of high quality electron beams in very high electric fields, moreover it has opened the possibility to design and conceive the compact setup \cite{Couprie2016}.
% FJ: One should say something on LWFA electron energies and mention and cite a few concrete examples. Not only on highest (GeV by capillary), but also $<$100 MeV.

Alternatively, Thomson scattering is capable to generate attosecond pulses from low-energy electrons (see e.g. \cite{Paz-2012}). Single-cycle attosecond pulse generation by Thomson scattering is possible using single-cycle laser pulses. However, single-cycle laser pulse generation is a big challenge in the visible spectral range. Therefore, alternative methods can be of significant interest.

In this paper, we propose a setup to generate waveform-controlled single-cycle attosecond pulses by Thomson scattering of intense THz pulses. In the terahertz (THz) spectral range, single-cycle pulse generation is straightforward \cite{Hoffmann2011,Hirori2011,Vicario2014,Fulop2014,Oh2014}. The peak electric field of the most intense THz pulses today reaches 40 MV/cm \cite{Vicario2014}, and there are many suggestions to increase the efficiency of THz generation \cite{Fulop-2016,Ofori-Okai2016,Palfalvi2016,Palfalvi2017,Vicario2015}. Here, we show that such THz pulses are a useful tool in generating single-cycle or waveform controlled attosecond pulses by Thomson scattering on ~30--40 MeV electrons, conveniently delivered by present laser-driven electron sources.

The paper is organized as follows. Section 2 describes the basics of Thomson scattering. Section 3 introduces the setup proposed in this work. Section 4 describes the calculation method. The results and their discussion are presented in Section 5. The coclusion is drawn in the last section.

\section{Scheme of the attosecond source}

\begin{figure*}[!ht]
\fbox{\includegraphics[width=\textwidth]{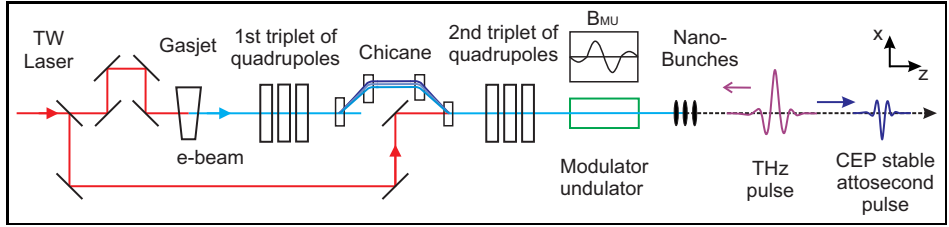}}
\caption{\label{fig:scheme} Scheme of the proposed LPWA-based, THz-driven Thomson scattering setup for singlecycle attosecond pulse generation.}
\end{figure*}

The proposed setup for the generation of single-cycle attosecond pulses is shown in Fig \ref{fig:scheme}. Relativistic electrons are generated in a LPWA. The electron beam is propagated through the first triplet of quadrupoles to reduce the divergence of the beam. However, the typical energy spread of a LPWA beam is of a few percent. Importantly, the slice energy spread can be reduced by a chicane. The second triplet of quadrupoles focuses the electron beam to the focus of the THz radiation. In order to generate a spatially periodic energy modulation of the electrons, a high-power laser pulse is superimposed on them in the modulator undulator (MU). This energy modulation leads to the formation of ultrathin ($<$20 nm) electron layers (nanobunches) in the drift space behind the MU. A counterpropagating strong-field THz pulse is focused to the position where the nanobunch has the smallest longitudinal size, where the electrons interact with the THz field.

Advantageously, a single laser system is used to drive the attosecond source, including the electron source, the nanobunching, and the THz source. A short-pulse pumped high-power few-cycle optical parametric chirped-pulse amplification (OPCPA) system can be an ideally fitting laser architecture \cite{Fulop-2007,Major-2009,Fattahi2014}. Whereas the 10 to 100-TW power OPCPA output pulses are suitable to drive the LPWA and the nanobunching, the sub-ps to few-ps pump pulses can be used to efficiently drive the THz source.

\section{Simulation methods}
The simulation consists of two main parts. Firstly, we calculated a LPWA process, the transport and manipulation of the electron bunch, including nanobunch generation. Secondly, Thomson-scattering was calculated with the initial electron bunch parameters from the first part.
%In this part, the initial parameters was the results of laser plasma wakefield acceleration. 

%\subsection{Laser Plasma wakefield accaleration}
%In laser plasma accelerator, the longitudinal Langmuir wave is excited by the laser which passes through the plasma. The perturbed electron density satisfies the equation
%\begin{equation}
%\left(\frac{\partial^2}{\partial t^2}+\omega_p^2\right)\frac{\delta n}{n_0}=c^2\nabla^2\frac{a_0^2}{2},
%\label{EDS_eq}
%\end{equation}
%where $\delta n$ is the perturbed electron density, $n_0$ is the unperturbed electron density, is the plasma frequency, and $c$ is the light velocity. The perturbed electron density excites the plasma wave potential which forms the accelerating field as in conventional accelerator. According to the laser strength, the laser plasma accelerator can be divided into three regimes: the linear regime, the quasilinear regime, and the nonlinear regime. In order to achieve monoenergetic, high charge electron beam, we considered here the nonlinear case of the bubble/blow-out regime, $a_0\gg 1$. The bubble mechanism works in this regime and can produce the quasi-monoenergetic electron beam. 

\subsection{Generation and manipulation of relativistic electrons}

In order to create a coherent attosecond source, a good quality electron beam with high charge and few tens of MeV energy is needed. According to the scaling law predicted by Gordienko and Pukhov \cite{Gordienko2005}, a high quality electron beam can be generated in the bubble regime by utilizing laser pulses with 800 nm central wavelength, a pulse duration of 8 fs, and 80 mJ pulse energy, corresponding to 10 TW power. Such a laser pulse can be focused in gas target to an optimized spot size of 3.7 $\mu$m corresponding to a peak intensity of $2.9\times 10^{19}$ W/cm$^2$. As per scaling the optimum density range for efficient electron acceleration in bubble regime can be estimated as $1.5\times 10^{19}$ cm$^{-3}$ . Following the mentioned scaling law \cite{Gordienko2005}, an electron beam of 34 MeV energy and 442 pC charge can be obtained in an acceleration distance of 78 $\mu$m. The main parameters of the used electron beam are listed in Table~\ref{tab:1}.

\begin{table}[htbp]
\centering
\caption{\label{tab:1}\bf Parameters of the electron beam from the LPWA.}
\begin{tabular}{ccc}
\hline
Parameter & Value  \\
\hline
Energy ($\gamma_0$) &68 \\
Energy spread ($\sigma_{\gamma_0}$) & 2~\% \\
Normalized emittance ($\gamma_0\varepsilon_{x,y}$) & 0.078 mm mrad \\
Transversal size ($\sigma_{x0}=\sigma_{y0}$)& 5~$\mu$m \\
Length ($\sigma_{z0}$) & 5~$\mu$m\\
Charge & 442 pC\\
\hline
\end{tabular}
\end{table}

The General Particle Tracer (GPT) numerical code was used for the simulation of the electron beam transport and the ultrathin electron layer generation. One million macroparticles were used in the calculation, whereby one macroparticle consisted of 2760 electrons. The electron beam from a LPWA typically has a large transversal divergence, the reduction of which is important for further use. Furthermore, it was shown that the generation of coherent radiation is more efficient, when the transversal size of the nanobunch is smaller \cite{TibaiPRL}. Therefore, the electron beam was focused by two quadrupole triplets to the THz beam waist. The gradients of the quadrupole triplets were optimized with a self-developed numerical code. In the best case 15~$\mu$m and 25~$\mu$m transversal sizes were calculated in the $x$ and $y$ directions, respectively. The first chicane decompresses the electron bunch longitudinally, and after the chicane the slice energy spread of the electron bunch is reduced from 2~\% to 0.2~\%. The main parameters of the quadrupoles and the chicane are listed in Table \ref{tab:2}.

\begin{table}[htbp]
\centering
\caption{\label{tab:2}\bf Parameters of the quadrupoles and the chicane.}
\begin{tabular}{p{5cm}p{2cm}}%{ccc}
\hline
Parameter & Value  \\
\hline
Distance between the gas jet and the 1st quadrupole tripled & 15 cm \\
1st quadrupole triplet gradients & 10.1/-7.1/3.2 \\
1st quadrupole triplet elements length & 10 cm \\
Chicane dipole strength & 0.043~T \\
Chicane dipole length & 15~cm \\
2nd quadrupole triplet gradients &-2.5/5.6/-6.2 \\
2nd quadrupole triplet elements length & 10 cm \\
\hline
\end{tabular}
\end{table}

\begin{figure}[htbp]
\centering
\fbox{\includegraphics[width=8 cm]{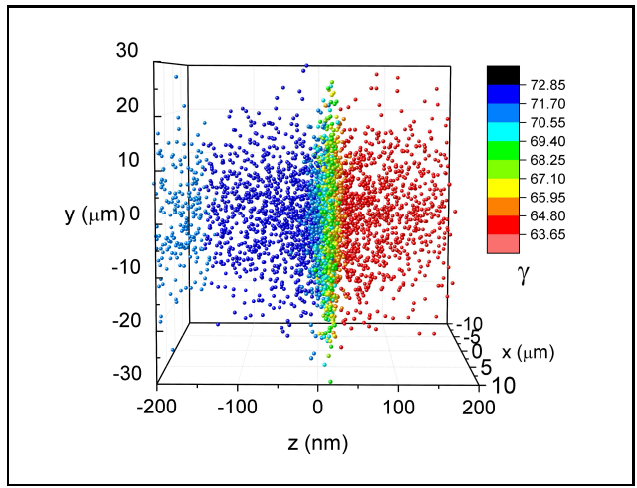}}
\caption{The spatial and energy distribution of macroparticles in our model. One sphere represents 2760 electrons. The energy distribution of macroparticles (together the real electrons) show the color scale.}
\label{fig:2}
\end{figure}

The relativistic electron beam is sent through the modulator undulator (MU), where a 17-TW power laser beam of 800 nm central wavelength is superimposed on it, in order to generate nanobunches. The MU used here has two periods. The magnetic field of the MU is trimmed in antisymmetric design ($1/4, -3/4, 3/4, -1/4$) along the electron propagation direction. After modulating the electron energy with the laser, a series of nanobunches is being formed. The individual nanobunches are separated by the modulator laser wavelength. The shortest nanobunch with only 16 nm length, containing a charge of 0.6~pC, was used in the simulation of Thomson scattering. The modulator laser and MU parameters are listed in Table \ref{tab:3}. The spatial distribution of the nanobunch is shown in Fig. \ref{fig:2}, where the color represents the energies of the macroparticles.

\begin{table}[htbp]
\centering
\caption{\label{tab:3}\bf Parameters of the modulator laser and the modulator undulator.}
\begin{tabular}{ccc}
\hline
Parameter & Value  \\
\hline
Laser wavelength ($\lambda_l$) & 800~nm \\
Laser peak power & 17 TW \\
Laser beam waist & 1 mm \\
Laser beam Rayleight length & 3.9 m \\
MU undulator parameter ($K_{MU}$)& 0.5 \\
MU period length ($\lambda_{MU}$) & 6.7 mm\\
\hline
\end{tabular}
\end{table}

\subsection{Thomson scattering}

\begin{figure}[htbp]
\centering
\fbox{\includegraphics[width=8 cm]{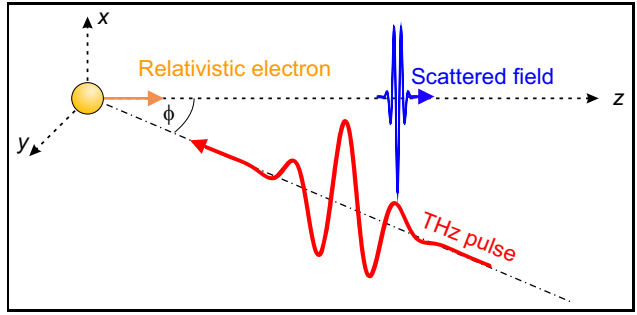}}
\caption{Scheme of Thomson scattering of a single-cycle THz pulse on relativistic electrons. Both the THz pulse as well as the scattered pulse are polarized along the $x$ direction. The incoming THz pulse propagates in the $yz$ plane.}
\label{fig:Thomson_basic}
\end{figure}

The interaction of a thin layer of relativistic electrons with a counterpropagating intense THz pulse results in Thomson scattering, schematically shown in Fig. \ref{fig:Thomson_basic}. During the interaction, the electromagnetic field of the THz pulse affects the movement of the electrons according to the Lorentz-law. The resulting oscillatory electron motion generates a radiation field, which can be determined based on Lienard-Wiecherd potentials \cite{Jackson} as:
\begin{equation}
\vec E(\vec r, t) = \left[ \frac{q_e\mu_0}{4\pi}\frac{\vec R\times\left(\left(\vec R-R\beta\right)\times\frac{\mathrm{d}}{\mathrm{d}t}\vec v\right)}{\left(R-\vec R\vec\beta\right)^3}\right]_{ret},
\label{eq:radiation_field}
\end{equation}
where $q_e$ is the charge of the electron, $\mu_0$ is the vacuum permeability, $\vec R$ is the vector pointing from the position of the electron at the retarded moment to the observation point, $\vec v$ is the velocity of the electron, $\vec \beta = \vec v/c$, and $c$ is the speed of light.

When the THz pulse with $\lambda_0$ central wavelength interacts with a relativistic electron, the  wavelength of the backscattered radiation field becomes \cite{Debus2010}
\begin{equation}
\lambda_r = \frac{1+a_0^2/2+\gamma^2\theta^2}{2\left(1+\cos\phi\right)}\cdot\frac{\lambda_0}{\gamma^2},
\label{eq:radiation_wavelength}
\end{equation}
where $\gamma=\left({1-^{v^2}/_{c^2}}\right)^{-^1/_2}$ is the relativistic factor of the electron, $\theta$ is the angle between the electron propagation direction and the observation axis, $\phi$ is the interaction angle between the THz pulse and the electron, and $a_0=q_eE_0/m_ec\omega_0$ is the normalized THz field strength, where $E_0$ is the peak electric field, $\omega_0=2\pi c/\lambda_0$ is the angular frequency, and $m_e$ is the mass of the electron.

The scattered field was determined according to Eq. \ref{eq:radiation_field}, whereby macroparticles, rather than single electrons, were taken into account. Accordingly, $q_e$ and $m_e$ was equal to the charge and the mass of one macroparticle, respectively. %%T Gy. az előző modtatot átírta
In order to use Eq. \ref{eq:radiation_field}, first it was necessary to solve the equation of motion for the macroparticles in the electromagnetic field of the THz pulse. The equation of motion reads as:
\begin{equation}
\frac{\mathrm{d}\vec{p}}{\mathrm{d}t}=q_0\left(\vec E(\rho,z,t)+\vec v(t)\times\vec B(\rho,z,t)\right),
\label{movement_eq}
\end{equation}
where $\vec p=m_0\gamma \vec v$ is the electron momentum vector, $m_0$ is the mass of the macroparticle, $q_0$ is the charge of the macroparticle, %T Gy. átírta%
$\rho$ is the radial coordinate, $t$ is the time. The electric and the magnetic fields of the THz pulse are given as follows, respectively:
\begin{eqnarray}
&&\vec E(\rho,z,t) = E_0\vec x \frac{w_0}{w(z)}\exp\left(-\frac{\rho^2}{w^2(z)}\right)\times \nonumber \\ 
&&\exp\left(-2\ln(2)\frac{\left(z/c+t\right)^2}{\tau^2}\right)\cos\left(k_0\left(z+ct\right)-\Psi(z)\right),
\label{efield_eq}
\end{eqnarray}
\begin{equation}
\vec B(\rho,z,t) = \frac{1}{k_0 c} \vec E(\rho,z,t)\times \vec k_0.
\label{bfield_eq}
\end{equation}
In these equations, $\vec x$ is the direction of the electric field polarization, $w_0$ is the THz beam waist radius, $w(z)=w_0\sqrt{1+\left(\frac{z}{z_R}\right)^2}$ is the beam radius at position $z$, $z_R=\frac{\pi w_0^2}{\lambda}$ is the Rayleigh length of the Gaussian beam, $\lambda_0$ is the central wavelength of the THz pulse, $\tau$ is the full width at half maximum (FWHM) of the intensity in time, $\vec k_0 = \frac{2\pi}{\lambda_0}\vec e_z$ is the wave vector,$\Psi(z) = \arctan{\left(\frac{z}{z_R}\right)}$ is the Gouy phase. Because single-cycle THz pulses were supposed, the FWHM of the pulses were $\tau = \frac{2\pi}{k_0c}$.

The energy of the electromagnetic pulse in the vacuum can be calculated according to $W = \int_{-\infty}^{\infty}\int_0^\pi\int_0^\infty \frac{1}{2}\epsilon_0 c \left|\vec E(\rho, z, t)\right|^2\rho\mathrm{d}\rho\mathrm{d}\phi\mathrm{d}t$ \cite{Diels_Rudolph}, so the field amplitude in Eq. (\ref{efield_eq}) was calculated as:
\begin{equation}
E_0 = \frac{2\sqrt{2}\left(\ln 2\right)^\frac{1}{4}\sqrt{W}}{\sqrt{\epsilon_0 c \pi^\frac{3}{2}\tau}w_0},
\label{eq:E_0}
\end{equation}
where $\epsilon_0$ is the vacuum permittivity.

For linear Thomson scattering, the total output radiation yield is proportional to the intensity of the THz pulse \cite{Debus2010, Esarey1993}. Therefore, the laser pulse must be focused to increase its intensity. In the calculations, $w_0 = c/\nu_0$ was assumed, where $\nu_0$ is the central frequency of the THz pulse. Using this focusing and the single-cycle conditions, Eq. \ref{eq:E_0} gives the following  energy- and frequency-dependent relationship:
\begin{equation}
E_0 = \frac{2\sqrt{2}\left(\ln 2\right)^\frac{1}{4}}{c\sqrt{\epsilon_0 c \pi^\frac{3}{2}}}\sqrt{W\nu_0^3},
\label{eq:E_0_2}
\end{equation}
We note that in the calculations of the next Section, head-on collision between the electron and the THz beams was assumed and Eq. \ref{eq:radiation_wavelength} can be simplified to:
\begin{equation}
\lambda_r = \left(1+a_0^2/2+\gamma^2\theta^2\right)\cdot\frac{\lambda_0}{4\gamma^2}.
\label{eq:simply_radiation_wavelength}
\end{equation}

%%%%%%%%%%%%%%%%%%%%%%%%%%%%%%%%%%%%%%%%%%%%%%%%%%%%%%%%%%%%%%%%%%%%%%%%%%%%%%%%%%%%%
%%%%%%%%%								RES AND DISC						%%%%%%%%%
%%%%%%%%%%%%%%%%%%%%%%%%%%%%%%%%%%%%%%%%%%%%%%%%%%%%%%%%%%%%%%%%%%%%%%%%%%%%%%%%%%%%%
\section{Results and discussion}

\begin{figure}[htbp]
\centering
\fbox{\includegraphics[width=8 cm]{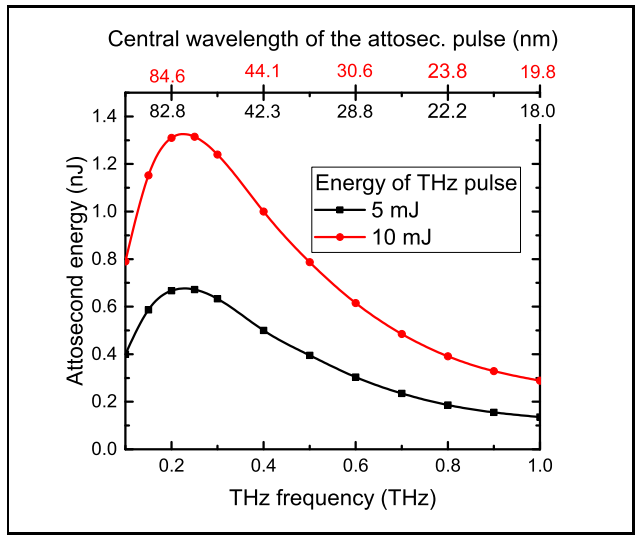}}
\caption{Energy of the generated single-cycle attosecond pulse as function of the central frequency of the THz pulse in case of 5 mJ (black curve) and 10 mJ (red curve) THz pulse energies. The upper axis shows the predicted central wavelength of the emitted radiation according to Eq. \ref{eq:radiation_wavelength} Values in black (red) correspond to 5 mJ (10 mJ).}
\label{fig:energy}
\end{figure}

We investigated the dependence of the scattered field energy on the frequency and energy of the THz pulse. The results are shown in Fig. \ref{fig:energy}. In accordance with previous theoretical predictions for laser pulses \cite{Debus2010,Esarey1993}, the radiated energy is proportional to the THz pulse energy. Therefore, the scattered radiation energy can be increased by increasing the energy of the THz pulse.

Figure \ref{fig:energy} shows that the radiated energy, as function of the THz frequency, has a maximum at about 0.2 THz. The initial increase at lower THz frequencies can be explained as follows. The number of radiated photons per electron in case of a single-cycle THz pulse is given by $N \approx \alpha \cdot a_0^2$ \cite{Ride1995}, where $\alpha$ is the fine structure constant. The THz pulse energy was kept constant along each curve in Fig. \ref{fig:energy}, whereby a single-cycle waveform and a diffraction-limited focal spot size was assumed at each value of the THz central frequency. The resulting $a_0^2 \propto W \times \nu_0$ scaling leads to $N \propto \nu_0$. In addition, the energy of the radiated photon also increases with the THz frequency (see Eq. \ref{eq:radiation_wavelength}). It means that the energy emitted by one electron increases with increasing $\nu_0$. As long as transversal effects can be neglected (see below), this leads to an increasing total emitted energy for an extended nanobunch.

However, as the THz frequency, and consequently the radiation frequency, increases, transversal effects for the extended electron sheet become significant. As we showed in \cite{TibaiPRL}, owing to coherent superposition, the solid angle of the emitted radiation for a nanobunch with a finite transversal size is smaller than that for a single electron. This also reduces the total radiated energy. Obviously, this effect becomes stronger for higher frequency (smaller wavelength) of the emitted radiation. This is the reason for the decreasing radiated energy with increasing THz frequency.

For a given THz frequency, the energy of the THz pulse is proportional to $a_0^2$. Therefore, according to Eq. \ref{eq:radiation_wavelength}, increasing the THz energy at a constant THz frequency results in increasing radiation wavelength as $\lambda_r \propto 1 + a_0^2/2$ for $\theta = 0$. This, together with the $a_0^2 \propto W \times \nu_0$ scaling mentioned above, causes a moderate variation of the radiation wavelength with the THz energy. By increasing the THz energy from 5 mJ to 10 mJ leads to about 2\% (10\%) shift in the radiation wavelength at 0.1 THz (1 THz) frequency.

% In the 0.1--1 THz frequency range considered here, $a_0$ varies between 0.15 and 0.47 (0.21 and 0.66) for $W = 5$ mJ (10 mJ). Consequently, the relative variation of the radiated wavelength  with the THz pulse energy remains less than 10\%.

%T. Gy.: Ha valamilyen lambda_r központi hullámhosszúságú sugárzást akarunk előállítani, akkor annak energiáját nem lehet pusztán a THz energiájával skálázni, mert nagyobb THz-es energia esetén hiába marad a THz frekvenciája konstans, a sugárzásé eltolódik. Ennek mértékét a következő bekezdés fogalmazza meg. Kis frekvenciákon 1-2 % az eltérés, 1 THz esetén pedig közel 10%

%Amikor a_0 1 közelébe ér valószínűleg telítődésbe megy az energianövekedés, ahogy azt az undulátorok esetén is megszoktuk, de az már 20 mJ feletti energiáknál lesz.

\begin{figure}
\centering
\fbox{\includegraphics[width= 8 cm]{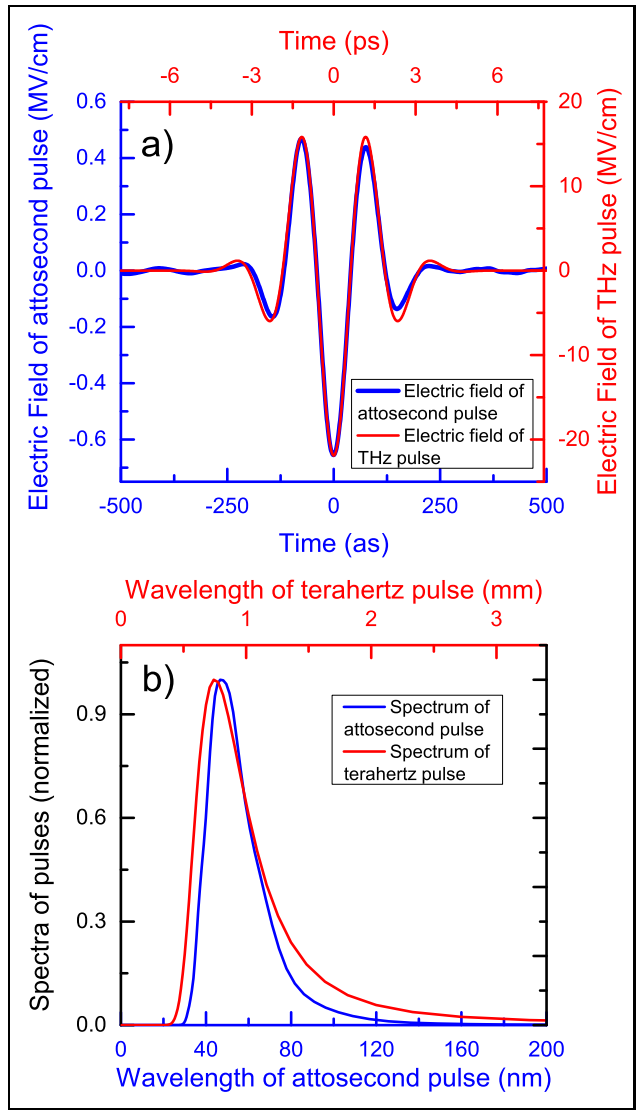}}
\caption{(a) The temporal waveforms of the Thomson-scattered pulse (blue) and the incoming THz pulse (red). (b) Spectra of the scattered field (blue) and the THz pulse (red).}
\label{fig:shape}
\end{figure}

A comparison of incoming THz and scattered attosecond pulse waveforms and spectra are shown in Fig. \ref{fig:shape}.a and \ref{fig:shape}.b, respectively. In Fig. \ref{fig:shape}.a the red curve shows a single-cycle THz pulse with 10 mJ energy and 0.4 THz central frequency. The blue curve shows the scattered field. The two waveforms are very similar. The pulse duration of the scattered pulse is less than 150 as with 48 nm central wavelength, consistent with the single-cycle waveform. The central frequency of the scattered field is about 17,000 times higher than the THz frequency, in good agreement with Eq. \ref{eq:radiation_wavelength}. This is clearly shown in Fig. \ref{fig:shape}.b, where the blue curve shows the spectrum of the attosecond pulse and the red curve shows the spectrum of the THz pulse on a 17,000 times larger wavelength scale.

We note that varying the interaction angle $\phi$ between the propagation direction of the electrons and the propagation direction of the THz pulse (see Fig. \ref{fig:Thomson_basic}) allows to use an electron bunch with still higher energy. For example, $\phi = 90^\circ$ allows for two times larger electron energy to  be used. Because the radiated energy is proportional to the square of the electron energy, up to four times higher emitted pulse energy can be expected.

% in addition to keeping the radiation coherent

Furthermore, a tunable narrowband extreme ultraviolet source can also be constructed based on the proposed method, whereby a tunable THz source can be used. A suitable THz source can be a semiconductor contact-grating device \cite{Fulop-2016}, pumped by the periodically intensity-modulated pulses from a dual-chirped optical parametric amplifier \cite{Toth2017}.

\section{Conclusion}

A novel scheme of single-cycle attosecond pulse generation was introduced, and investigated by numerical simulations. The entirely laser-based scheme uses electrons of a few tens-of-MeV energy generated by a laser wakefield accelerator, subsequent nanobunching to ultrathin electron layers in a modulating undulator driven by TW laser pulses, and the Thomson scattering of intense THz pulses to generate nJ-level attosecond pulses in the few tens of nm wavelength range. The scheme can be driven for example by a short-pulse pumped few-cycle OPCPA system, whereby the strong THz pulses can be conveniently generated by the pump laser. The waveform of the generated attosecond pulses closely resembles that of the THz pulses and can be flexibly shaped by shaping the THz pulse.

% Bibliography
\bibliography{sample}{}

\begin{thebibliography}{10}

\bibitem{Krausz2009}
Ferenc Krausz and Misha Ivanov.
\newblock Attosecond physics.
\newblock {\em Rev. Mod. Phys.}, 81:163--234, Feb 2009.

\bibitem{Zholents2004}
Alexander~A Zholents and William~M Fawley.
\newblock Proposal for intense attosecond radiation from an x-ray free-electron
  laser.
\newblock {\em Phys. Rev. Lett.}, 92(22):224801, 2004.

\bibitem{Saldin2006}
Evgenij~L Saldin, Evgeny~A Schneidmiller, and Mikhail~V Yurkov.
\newblock Self-amplified spontaneous emission fel with energy-chirped electron
  beam and its application for generation of attosecond x-ray pulses.
\newblock {\em Phys. Rev. ST Accel. Beams}, 9(5):050702, 2006.

\bibitem{Marinelli2013}
A.~Marinelli, E.~Hemsing, and J.~B. Rosenzweig.
\newblock Using the relativistic two-stream instability for the generation of
  soft-x-ray attosecond radiation pulses.
\newblock {\em Phys. Rev. Lett.}, 110:064804, Feb 2013.

\bibitem{Dunning2013}
D.~J. Dunning, B.~W.~J. McNeil, and N.~R. Thompson.
\newblock Few-cycle pulse generation in an x-ray free-electron laser.
\newblock {\em Phys. Rev. Lett.}, 110:104801, Mar 2013.

\bibitem{Wu2011}
H.-C. Wu, J.~Meyer-ter Vehn, B.~M. Hegelich, and J.~C. Fern\'andez.
\newblock Nonlinear coherent thomson scattering from relativistic electron
  sheets as a means to produce isolated ultrabright attosecond x-ray pulses.
\newblock {\em Phys. Rev. ST Accel. Beams}, 14:070702, Jul 2011.

\bibitem{Paz-2012}
A.~Paz, S.~Kuschel, C.~Rödel, M.~Schnell, O.~Jäckel, M.~C. Kaluza, and G.~G.
  Paulus.
\newblock Thomson backscattering from laser-generated, relativistically moving
  high-density electron layers.
\newblock {\em New J. Phys.}, 14:093018, 2012.

\bibitem{Lu02013}
W.~Luo, H.~B. Zhuo, Y.~Y. Ma, Y.~M. Song, Z.~C. Zhu, T.~P. Yu, and M.~Y. Yu.
\newblock Attosecond thomson-scattering x-ray source driven by laser-based
  electron acceleration.
\newblock {\em Appl. Phys. Lett.}, 103(17):174103, 2013.

\bibitem{Hack2017}
Szabolcs Hack, Sándor Varró, and Attila Czirják.
\newblock Carrier-envelope phase controlled isolated attosecond pulses in the
  nm wavelength range, based on superradiant nonlinear thomson-backscattering.
\newblock {\em ArXiv e-prints}, September 2017.

\bibitem{Sansone2006}
G.~Sansone, E.~Benedetti, F.~Calegari, C.~Vozzi, L.~Avaldi, R.~Flammini,
  L.~Poletto, P.~Villoresi, C.~Altucci, R.~Velotta, S.~Stagira,
  S.~De~Silvestri, and M.~Nisoli.
\newblock Isolated single-cycle attosecond pulses.
\newblock {\em Science}, 314(5798):443--446, 2006.

\bibitem{Ma-2014}
W.~J. Ma, J.~H. Bin, H.~Y. Wang, M.~Yeung, C.~Kreuzer, M.~Streeter, P.~S.
  Foster, S.~Cousens, D.~Kiefer, B.~Dromey, X.~Q. Yan, J.~Meyer ter Vehn,
  M.~Zepf, and J.~Schreiber.
\newblock Bright subcycle extreme ultraviolet bursts from a single dense
  relativistic electron sheet.
\newblock {\em Phys. Rev. Lett.}, 113:235002, 2014.

\bibitem{Tanaka2014}
T.~Tanaka.
\newblock Proposal to generate an isolated monocycle x-ray pulse by
  counteracting the slippage effect is free-electron lasers.
\newblock {\em Phys. Rev. Lett.}, 114:044801, Jan 2015.

\bibitem{TibaiPRL}
Z.~Tibai, Gy. Tóth, M.~I. Mechler, J.~A. Fülöp, G.~Almási, and J.~Hebling.
\newblock Proposal for carrier-envelope-phase stable single-cycle attosecond
  pulse generation in the extreme-ultraviolet range.
\newblock {\em Phys. Rev. Lett.}, 113:104801, Sep 2014.

\bibitem{TibaiarXive}
Zoltán Tibai, György Tóth, Zsuzsanna Nagy-Csiha, József~András Fülöp,
  Gábor almási, and János Hebling.
\newblock Investigation of the newly proposed carrier-envelope-phase stable
  attosecond pulse source.
\newblock {\em ArXiv e-prints}, April 2016.

\bibitem{Toth2016}
György Tóth, Zoltán Tibai, Zsuzsanna Nagy-Csiha, Zsuzsanna Márton, Gábor
  Almási, and János Hebling.
\newblock Investigation of novel shape-controlled linearly and circularly
  polarized attosecond pulse sources.
\newblock {\em Nucl. Instrum. Methods Phys. Res. B}, 369(Supplement C):2 -- 8,
  2016.
\newblock Photon and fast Ion induced Processes in Atoms, Molecules and
  Nanostructures (PIPAMON).

\bibitem{Leemans2006}
W.~P. Leemans, B.~Nagler, A.~J. Gonsalves, Cs. Tóth, K.~Nakamura, C.~G.~R.
  Geddes, E.~Esarey, C.~B. Schroeder, and S.~M. Hooker.
\newblock Gev electron beams from a centimetre-scale accelerator.
\newblock {\em Nat. Phys.}, 2:696--699, 2006.

\bibitem{Lundh-2011}
O~Lundh, J~Lim, C~Rechatin, L~Ammoura, A~Ben-Ismail, X~Davoine, G~Gallot, J-P
  Goddet, E~Lefebvre, V~Malka, and J~Faure.
\newblock Few femtosecond, few kiloampere electron bunch produced by a
  laser–plasma accelerator.
\newblock {\em Nat. Phys.}, 7:219–222, 2011.

\bibitem{Esarey2009}
E.~Esarey, C.~B. Schroeder, and W.~P. Leemans.
\newblock Physics of laser-driven plasma-based electron accelerators.
\newblock {\em Rev. Mod. Phys.}, 81:1229--1285, Aug 2009.

\bibitem{Couprie2016}
M.~E. Couprie, M.~Labat, C.~Evain, C.~Szwaj, S.~Bielawski, N.~Hubert,
  C.~Benabderrahmane, F.~Briquez, L.~Chapuis, F.~Marteau, M.~Valléau,
  O.~Marcouillé, P.~Marchand, M.~Diop, J.~L. Marlats, K.~Tavakoli, D.~Zerbib,
  L.~Cassinari, F.~Bouvet, C.~Herbeaux, C.~Bourassin-Bouchet, D.~Dennetière,
  F.~Polack, A.~Lestrade, M.~Khojoyan, W.~Yang, G.~Sharma, P.~Morin, and
  A.~Loulergue.
\newblock Strategies towards a compact xuv free electron laser adopted for the
  lunex5 project.
\newblock {\em Journal of Modern Optics}, 63(4):309--323, 2016.

\bibitem{Hoffmann2011}
M.~C. Hoffmann and J.~A. Fülöp.
\newblock Intense ultrashort terahertz pulses: generation and applications.
\newblock {\em J. Phys. D: Appl. Phys.}, 44:083001, 2011.

\bibitem{Hirori2011}
H.~Hirori, A.~Doi, F.~Blanchard, and K.~Tanaka.
\newblock Single-cycle terahertz pulses with amplitudes exceeding 1 mv/cm
  generated by optical rectification in linbo$_3$.
\newblock {\em Appl. Phys. Lett.}, 98:091106, 2011.

\bibitem{Vicario2014}
C.~Vicario, A.~V. Ovchinnikov, S.~I. Ashitkov, M.~B. Agranat, V.~E. Fortov, and
  C.~P. Hauri.
\newblock Generation of 0.9-mj thz pulses in dstms pumped by a cr:mg$_2$sio$_4$
  laser.
\newblock {\em Appl. Phys. Lett.}, 98:091106, 2011.

\bibitem{Fulop2014}
J.~A. Fülöp, Z.~Ollmann, C.~Lombosi, C.~Skrobol, S.~Klingebiel, L.~Pálfalvi,
  F.~Krausz, S.~Karsch, and J.Hebling.
\newblock Efficient generation of thz pulses with 0.4 mj energy.
\newblock {\em Opt. Express}, 22(17):20155--20163, 2014.

\bibitem{Oh2014}
T.~I. Oh, Y.~J. Yoo, Y.~S. You, and K.~Y. Kim.
\newblock Generation of strong terahertz fields exceeding 8 mv/cm at 1 khz and
  real-time beam profiling.
\newblock {\em Appl. Phys. Lett.}, 105(4):041103, 2014.

\bibitem{Fulop-2016}
J.~A. Fülöp, Gy. Polónyi, B.~Monoszlai, G.~Andriukaitis, T.~Balciunas,
  A.~Pugzlys, G.~Arthur, A.~Baltuska, and J.~Hebling.
\newblock Highly efficient scalable monolithic semiconductor terahertz pulse
  source.
\newblock {\em Optica}, 3:1075, 2016.

\bibitem{Ofori-Okai2016}
Benjamin~K. Ofori-Okai, Prasahnt Sivarajah, W.~Ronny Huang, and Keith~A.
  Nelson.
\newblock Thz generation using a reflective stair-step echelon.
\newblock {\em Opt. Express}, 24(5):5057--5068, Mar 2016.

\bibitem{Palfalvi2016}
László Pálfalvi, Zoltán Ollmann, Levente Tokodi, and János Hebling.
\newblock Hybrid tilted-pulse-front excitation scheme for efficient generation
  of high-energy terahertz pulses.
\newblock {\em Opt. Express}, 24(8):8156--8169, 2016.

\bibitem{Palfalvi2017}
László Pálfalvi, György Tóth, Levente Tokodi, Zsuzsanna Márton,
  József~András Fülöp, Gábor Almási, and János Hebling.
\newblock Numerical investigation of a scalable setup for efficient terahertz
  generation using a segmented tilted-pulse-front excitation.
\newblock {\em Opt. Express}, 25(23), 2017.

\bibitem{Vicario2015}
C.~Vicario, M.~Jazbinsek, A.~V. Ovchinnikov, O.~V. Chefonov, S.~I. Ashitkov,
  M.~B. Agranat, and C.~P. Hauri.
\newblock High efficiency thz generation in dstms, dast and oh1 pumped by
  cr:forsterite laser.
\newblock {\em Opt. Express}, 23(4):4573--4580, Feb 2015.

\bibitem{Fulop-2007}
J~A Fülöp, Zs~Major, A~Henig, S~Kruber, R~Weingartner, T~Clausnitzer, E-B
  Kley, A~Tünnermann, V~Pervak, A~Apolonski, J~Osterhoff, R~Hörlein,
  F~Krausz, and S~Karsch.
\newblock Short-pulse optical parametric chirped-pulse amplification for the
  generation of high-power few-cycle pulses.
\newblock {\em New J. Phys.}, 9:438, 2007.

\bibitem{Major-2009}
Zsuzsanna Major, Sergei~A. Trushin, Izhar Ahmad, Mathias Siebold, Christoph
  Wandt, Sandro Klingebiel, Tie-Jun Wang, József~András Fülöp, Andreas
  Henig, Sebastian Kruber, Raphael Weingartner, Antonia Popp, Jens Osterhoff,
  Rainer Hörlein, Joachim Hein, Vladimir Pervak, Alexander Apolonski, Ferenc
  Krausz, and Stefan Karsch.
\newblock Basic concepts and current status of the petawatt field synthesizer -
  a new approach to ultrahigh field generation.
\newblock {\em Rev. Laser Eng.}, 37(6):431--436, 2009.

\bibitem{Fattahi2014}
Hanieh Fattahi, Helena~G. Barros, Martin Gorjan, Thomas Nubbemeyer, Bidoor
  Alsaif, Catherine~Y. Teisset, Marcel Schultze, Stephan Prinz, Matthias
  Haefner, Moritz Ueffing, Ayman Alismail, L\'{e}n\'{a}rd V\'{a}mos, Alexander
  Schwarz, Oleg Pronin, Jonathan Brons, Xiao~Tao Geng, Gunnar Arisholm, Marcelo
  Ciappina, Vladislav~S. Yakovlev, Dong-Eon Kim, Abdallah~M. Azzeer, Nicholas
  Karpowicz, Dirk Sutter, Zsuzsanna Major, Thomas Metzger, and Ferenc Krausz.
\newblock Third-generation femtosecond technology.
\newblock {\em Optica}, 1(1):45--63, Jul 2014.

\bibitem{Gordienko2005}
S.~Gordienko and A.~Pukhov.
\newblock Scalings for ultrarelativistic laser plasmas and quasimonoenergetic
  electrons.
\newblock {\em Phys. Plasmas}, 12:043109, 2005.

\bibitem{Jackson}
J.~D. Jackson.
\newblock {\em Classical Electrodynamics}.
\newblock Wiley, 2007.

\bibitem{Debus2010}
A.~D. Debus, M.~Bussmann, M.~Siebold, A.~Jochmann, U.~Schramm, T.~E. Cowan, and
  R.~Sauerbrey.
\newblock Traveling-wave thomson scattering and optical undulators
  forhigh-yield euv and x-ray sources.
\newblock {\em Appl. Phys. B}, 100(1):61--76, Jul 2010.

\bibitem{Diels_Rudolph}
J.-C. Diels and W.~Rudolph.
\newblock {\em Ultrashort Laser Pulse Phenomena}.
\newblock Elsevier Inc., 2006.

\bibitem{Esarey1993}
Eric Esarey, Sally~K. Ride, and Phillip Sprangle.
\newblock Nonlinear thomson scattering of intense laser pulses from beams and
  plasmas.
\newblock {\em Phys. Rev. E}, 48(4):3003--3021, 1993.

\bibitem{Ride1995}
Sally~K. Ride, Eric Esarey, and Michael Baine.
\newblock Thomson scattering of intense lasers from electron beams at arbitrary
  interaction angles.
\newblock {\em Phys. Rev. E}, 52:5425--5442, Nov 1995.

\bibitem{Toth2017}
György Tóth, József~András Fülöp, and János Hebling.
\newblock Periodically intensity-modulated pulses by optical parametric
  amplification for multicycle tunable terahertz pulse generation.
\newblock {\em Opt. Express}, 25(23), 2017.

\end{thebibliography}
\bibliographystyle{unsrt}
%\printbibliography

%Manual citation list
%\begin{thebibliography}{1}
%\bibitem{Zhang:14}
%Y.~Zhang, S.~Qiao, L.~Sun, Q.~W. Shi, W.~Huang, %L.~Li, and Z.~Yang,
 % \enquote{Photoinduced active terahertz metamaterials with nanostructured
  %vanadium dioxide film deposited by sol-gel method,} Opt. Express \textbf{22},
  %11070--11078 (2014).
%\end{thebibliography}

\end{document}